\documentclass[prl,preprint,superscriptaddress,amsmath,amssymb]{revtex4}

\usepackage{amsmath}
\usepackage{rotating}
\usepackage{graphicx}
\usepackage{epstopdf}

\usepackage{eso-pic}
\usepackage{type1cm}
\makeatletter
  \AddToShipoutPicture{%
    \setlength{\@tempdimb}{0.5\paperwidth}%
    \setlength{\@tempdimc}{0.5\paperheight}%
    \setlength{\unitlength}{1pt}%
    \put(\strip@pt\@tempdimb,\strip@pt\@tempdimc){%

    }
}
\newcommand{\sgn}{\mathop{\mathrm{sgn}}}

\begin{document}

\title{Thermophoretically Driven Carbon Nanotube Oscillators}
\author{V. R. Coluci}\email[\footnotesize{Electronic address: }]{vitor@ft.unicamp.br}

\affiliation{School of Technology, 
             University of Campinas - UNICAMP 13484-332, Limeira,
             SP, Brazil}

\author{V. S. Tim\'oteo}
\affiliation{School of Technology, 
             University of Campinas - UNICAMP 13484-332, Limeira,
             SP, Brazil}

\author{D. S. Galv\~ao}
\affiliation{Applied Physics Department, Institute of Physics P.O.Box 6165,
             University of Campinas - UNICAMP 13083-970, Campinas,
             SP, Brazil}


\begin{abstract}
The behavior of a nanodevice based upon double-walled carbon nanotube oscillators driven by periodically applied thermal gradients (7 and 17 K/nm) is investigated by numerical calculations and classical molecular dynamics simulations. Our results indicate that thermophoresis can be effective to initiate the oscillator and that suitable heat pulses may provide an appropriate way to tune its behavior. Sustained regular oscillatory as well as chaotic motions were observed for the systems investigated in this work.
\end{abstract}



\maketitle

The remarkable properties of carbon nanotubes (CNTs) have allowed the proposition, design and exploitation of nanomechanical devices. One example is the CNT oscillator proposed by Zheng and Jiang \cite{zheng-prl} based upon the experimental realization of low-friction nanoscale linear bearings \cite{zettl}. Linear nanomotors based on coupled CNTs have also been experimentally realized \cite{nakayama1,nakayama2}. CNT oscillators, in their simplest form, consist of a double-walled CNT where the inner tube can move inside the outer tube in an oscillatory low-friction motion of high frequency. This motion is induced and maintained by van der Waals interactions between the inner and the outer tubes. Many works have contributed to understand the characteristics of CNT oscillators such as their oscillation frequency dependence upon CNT parameters \cite{zheng-prb}, stability \cite{vitor,groove,apl1}, friction \cite{rivera,servantie-1,tangney}, energy dissipation \cite{servantie-prl,guo,zhao,popov1}, and chaotic behavior \cite{vitor-caos}. Key issues regarding these devices are how to initialize them and how to sustain the motion in a controllable way. Different strategies have been proposed to start the oscillatory motion, e.g., by applying magnetic and electrical fields \cite{vitor,popov2}, by accelerating encapsulated charged elements located inside the inner tube \cite{ions}, and by thermal expansion of encapsulated gases \cite{gas}. 

In this letter we propose an alternative way to start, maintain, and control the oscillatory motion, based on thermophoresis, i.e., the use of thermal gradients to induce mass transport. In this case, the thermal gradient is imposed on the outer tube by external agents such as electrical currents. This approach was inspired on the recent findings that demonstrated the possibility of producing nanoscale thermal motors based on CNTs \cite{rurali}. Thermophoretic mass transport through CNTs has been theoretically investigated for solid gold nanoparticles \cite{schoen} and water nanodroplets \cite{zambrano}, but not yet for CNT-based oscillators. 

\begin{figure}
\begin{center}
\includegraphics[angle=0,scale=0.4]{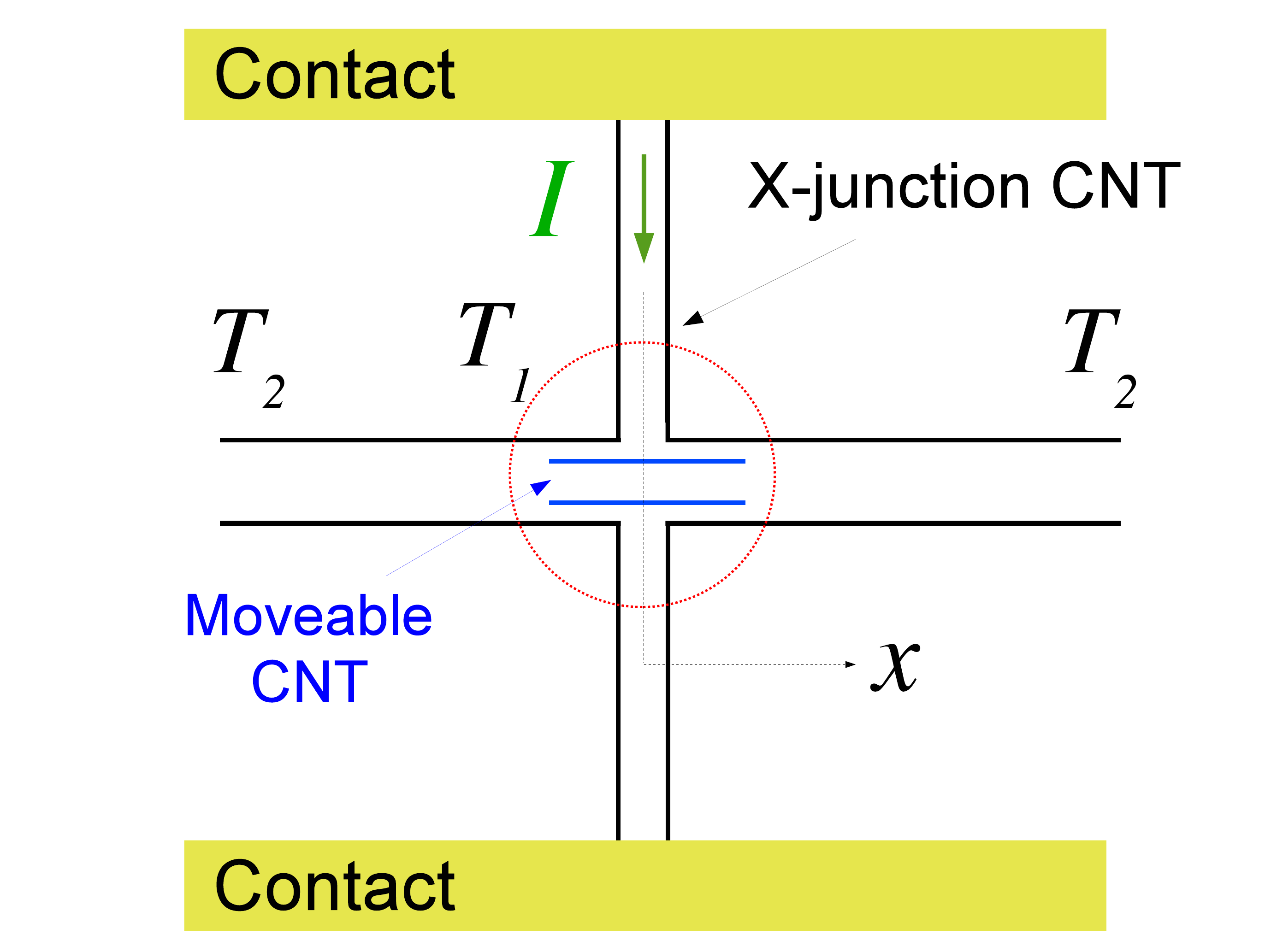}
\vspace{-0.4cm}
\caption{(Color online) Schematic representation of a nanodevice used to allow starting and controlling a CNT oscillator. The device is suspended between two contacts and it is composed by a X-like junction, which serves as support and also provides the outer tube of the oscillator, and an inner moveable tube.} 
\end{center}
\end{figure}

\vspace{-0.1 cm}

The concept of the thermophoretically driven device is presented in Fig.~1. A CNT containing a ``X''-like junction and a movable internal CNT is placed between two contacts. An electrical current $I$ can be injected through the contacts to create thermal gradients along the outer tube, with temperatures $T_1$ and $T_2$ ($T_1 > T_2$). The Joule heating will create a hotter region at the junction producing a heat flux along the $x$-direction. As will demonstrate below suitable choices of periodic electrical current pulses provide ways to start, tune and control the oscillatory behavior of the inner tube. It is important to mention that the configurations of tubes on the device of Fig.~1 are feasible to be obtained with current technological capabilities \cite{suspended-review}.

In order to investigate the dynamics of the device shown in Fig.~1 we have used analytical models and molecular dynamics (MD) simulations. For the analytical part we considered a simplified model where the motion of the inner tube of mass $m$, represented by the center of mass coordinate $x$, is described by the following equation of a non-linear, damped, driven oscillator
\begin{equation}
\displaystyle m\ddot{x} - F(x) + \gamma\sgn(\dot{x})|\dot{x}|^2 = \alpha \frac{dT}{dx}\; p(t).
\end{equation}
where $F$ is the restoring force due to van der Waals interactions between inner and outer tubes, the damped term is associated with a friction proportional to $|\dot{x}|^2$ (Ref.~11\nocite{tangney}), with a damped coefficient $\gamma$. The effect of the heating caused by the electrical current is translated into a driven ``thermal'' force which is proportional to the thermal gradient $dT/dx$. This model takes into account only the translational motion considering all tubes rigid, discarding rotational and vibrational modes. Despite the existence of these effects in real situations, the simplified analysis of the translational motion allows a fast estimative of the relative importance and contributions of geometrical (tube lengths and diameters, etc.) and dynamical aspects (temperature, damping constants, etc.) that can be obtained by numerical integration of Eq. (1). These analysis are combined with classical MD simulations based on the Brenner's potential \cite{brenner} (explicitly taking into account rotations and vibrations) to validate Eq. (1) and to extract the relevant parameters ($\gamma$, $\alpha$, and $|F|$). Non-bonded interactions were described by a Lennard-Jones 6-12 potential with $\sigma = 3.37$ {\AA} and $\epsilon = 4.2038$ meV \cite{mao,lj}. The equations of motion were integrated with a third-order Nordisieck predictor-corrector algorithm \cite{nordsieck} using a time step of 0.5 fs. We considered the (10,10)@(15,15) double-walled CNT as a model for the oscillator, with lengths of 100 nm and 5 nm for the outer and inner tubes, respectively \cite{mov}. The atoms within 0.5 nm of the outer tube extremities were kept fixed during the simulations. To create the thermal gradient on the outer tube, the Berendsen's thermostat \cite{berendsen} was applied to about 1100 atoms that were within 5 nm of each fixed extremity. We considered thermal gradients of 7 and 17 K/nm that are of the same order of magnitude that can be experimentally obtained \cite{zettl1}. Such large thermal gradient values also provided high signal-to-noise ratios in our MD simulations. Initially, we established a constant thermal gradient on the outer tube while the inner tube was separately equilibrated at 300 K. Then, we inserted the inner tube in the middle the outer tube and subtracted inner tube center of mass velocity from the velocity of its individual atoms. Fig.~2~(a) displays the time evolution of the center of mass of the inner tube and the resulting fitting curves, obtained with the following parameters: $|F|= 1.92 $ nN, $\gamma = 1.48~(2.24) $ pN ps$^2$/{\AA}$^2$, and $\alpha = 8.08~(6.26) $ pN nm/K for $7~(17)$ K/nm. Despite the simplicity of the model the fitting is quite good. Discrepancies are present and are due to other effects not included in the model such as the tube rotations and vibrations. The imposed thermal gradient causes an acceleration of the inner tube initially at rest. The acceleration is larger for larger thermal gradients. When reaching the outer tube extremity, the inner tube experiences a restoring force which causes its retraction. Due to the presence of a constant thermal gradient, the inner tube is accelerated towards the extremity of the outer tube. 

In order to simulate a temporal dependence of the driving force we used a periodic square wave-like pulse function $p(t)$ ($0\leq p(t)\leq 1$), which will be responsible to ``turn on'' (current turned on) and ``turn off'' (current turned off) the thermal gradient. The $dT/dx$ behavior as a function of $x$ is determined by the device geometry, electrical current pulse through $p(t)$, and by the thermal CNT conductivity values. Our MD simulations showed that a constant $dT/dx$ is obtained (after a transient period) for a large range of different temperatures \cite{mov}. These results also showed that thermal gradients can be a very effective way to initialize the oscillatory regime of the inner tube. 

In order to obtain $x(t)$, Eq. (1) was then numerically integrated using the fourth-order Runge-Kutta algorithm with a fixed time step of 1 fs, with $x(0)=$ 1 {\AA} and $\dot{x}(0)=$ 0. Two examples of possible regimes are shown in Fig.~2~(b). When the thermal force is time independent ($p(t)=1$, Fig.~2(b), curve~1), the inner tube is initially accelerated ($x(t)\propto t^2$) until reaching one of the extremity of the outer tube where it is subjected to the restoring force $F$. This force leads the inner tube to retract until some point where $F$ is compensated by the thermal force which sends it back to the extremity. This process continues back and forth until the inner tube is stopped. This is analog to the case of letting a steel ball to fall towards a rigid ground subject to the action of the gravitational and frictional forces. On the other hand, when a periodic pulse is used, the behavior of the inner tube is completely different (Fig.~2~(b), curve~2). We can see that the pulse can initialize and maintain the inner tube oscillations. The example presented in Fig.~2~(b) is for a tube movement that is not regular and does not show a well defined oscillatory period. By tuning the period of thermal pulses regular and periodic oscillations can also be obtained, as shown below. 

\begin{figure}
\begin{center}
\includegraphics[angle=0,scale=0.5]{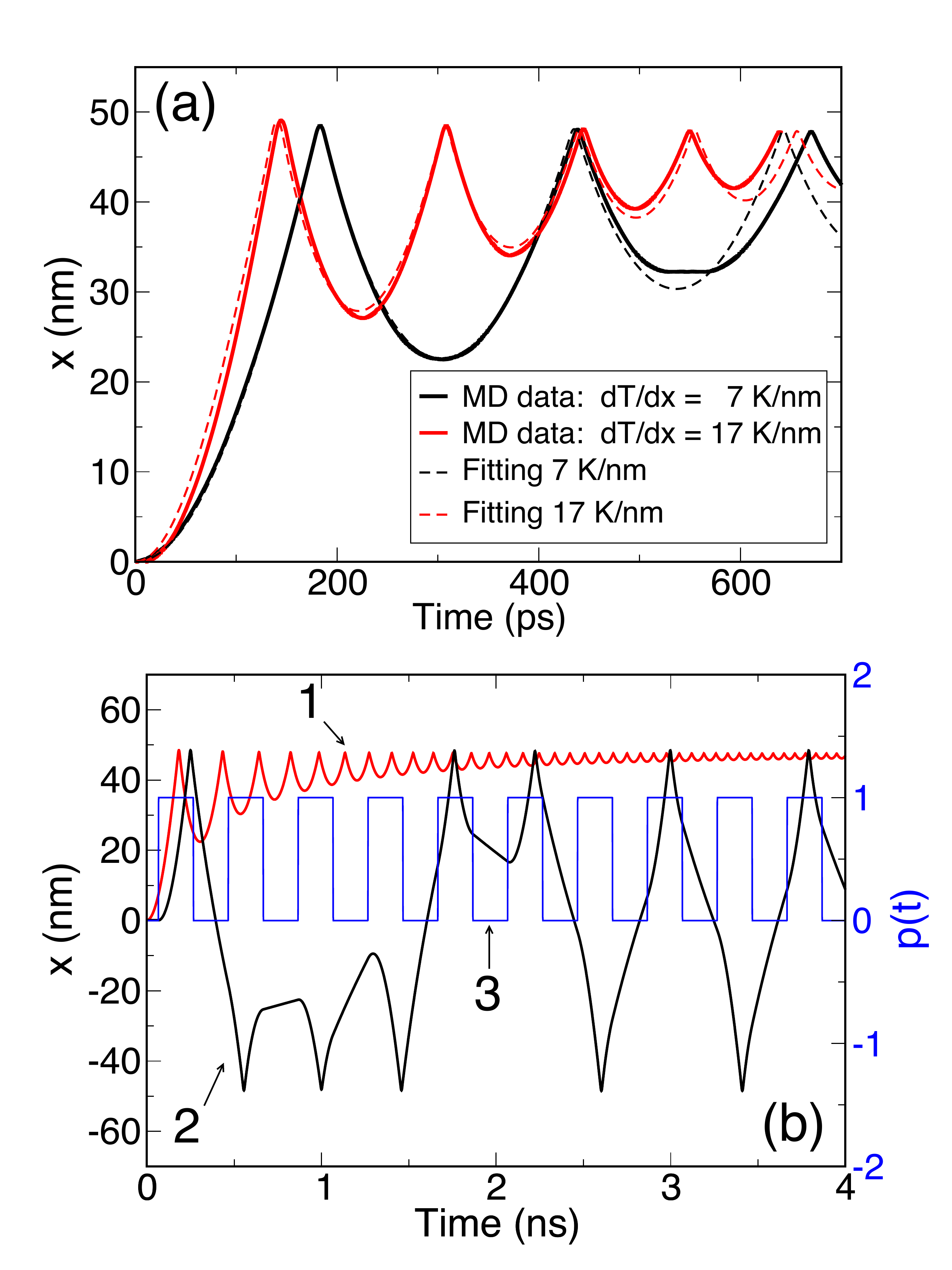}
\vspace{-0.4cm}
\caption{(Color online) (a) Time evolution of the inner tube obtained from MD simulations for thermal gradients of 7 K/nm and 17 K/nm. The fittings using Eq.~(1) are also presented. (b) Time evolution of the inner tube center of mass obtained from numerical integration of Eq.~(1) for two different situations with $dT/dx$ = 7 K/nm. Curve 1 represents the case where the thermal gradient is ``turned on'' in the beginning of the process and kept on this state, or equivalently, making $p(t)=1$. Curve 2 displays the case where a driven periodic pulse $p(t)$ with a period of 400 ps (represented by the curve 3) is used.}
\end{center}
\end{figure}

\begin{figure}
\begin{center}
\includegraphics[angle=0,scale=0.3]{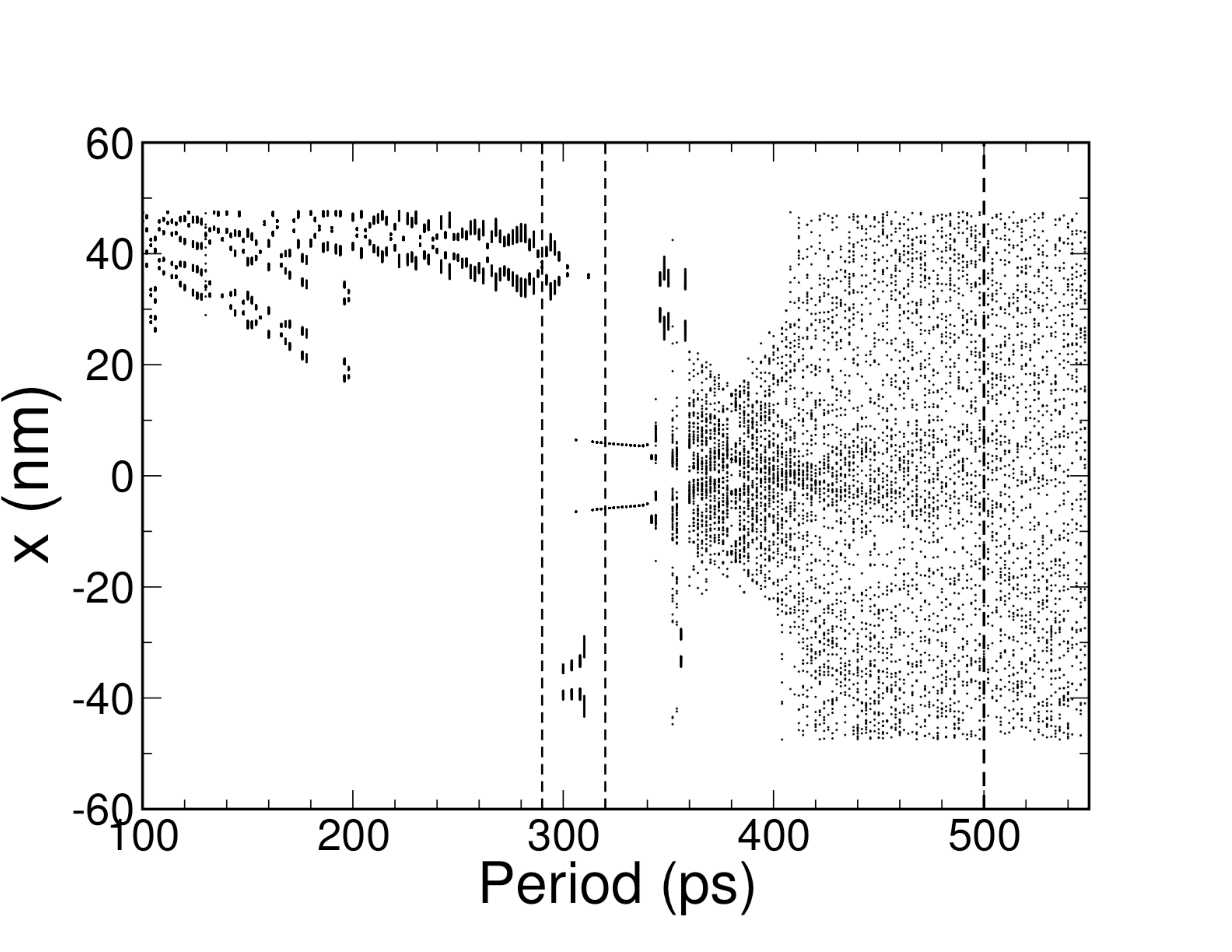}
\vspace{-0.5cm}
\caption{Bifurcation diagram for $x$ as a function of the period of $p(t)$. The data correspond to the same CNT oscillator and conditions of Fig.~2~(b).} 
\end{center}
\end{figure}

For a fixed value of $dT/dx$, the overall behavior of the inner tube for different values of the driven period of $p(t)$ can be analyzed using the bifurcation diagram shown in Fig.~3. The diagram presents the values of $x$ collected after 30 ns in a 60 ns-integration at times that are in phase with the period of $p(t)$. As we can see the diagram exhibits a very complex structure. However, it is possible to determine values of the period that provide sustained (i.e., regular oscillatory) and chaotic motions. These values are illustrated by the vertical dashed lines in Fig. 3 and their corresponding temporal profile movements are shown in Fig. 4. For a period of 290 ps, the inner tube reaches, after a transient period, a sustainable oscillatory motion on one side of the outer tube with the oscillatory period matching the imposed one. On the other hand, for the driven period of 320 ps the inner tube also reaches a sustainable motion but now its displacement covers both (right and left) sides of the outer tube. However, in this case, the oscillatory period is twice the driven one. This result is similar to what is observed in several non-linear dynamical systems, where the period-doubling is an indication of a route to chaos \cite{caos}. Finally, for a driven period of 500 ps, the inner tube again reaches both sides of the outer tube but the motion is chaotic, characterized by the large spreading of $x$ values shown in Fig. 3.

\begin{figure}[ht]
\begin{center}
\includegraphics[angle=0,scale=0.5]{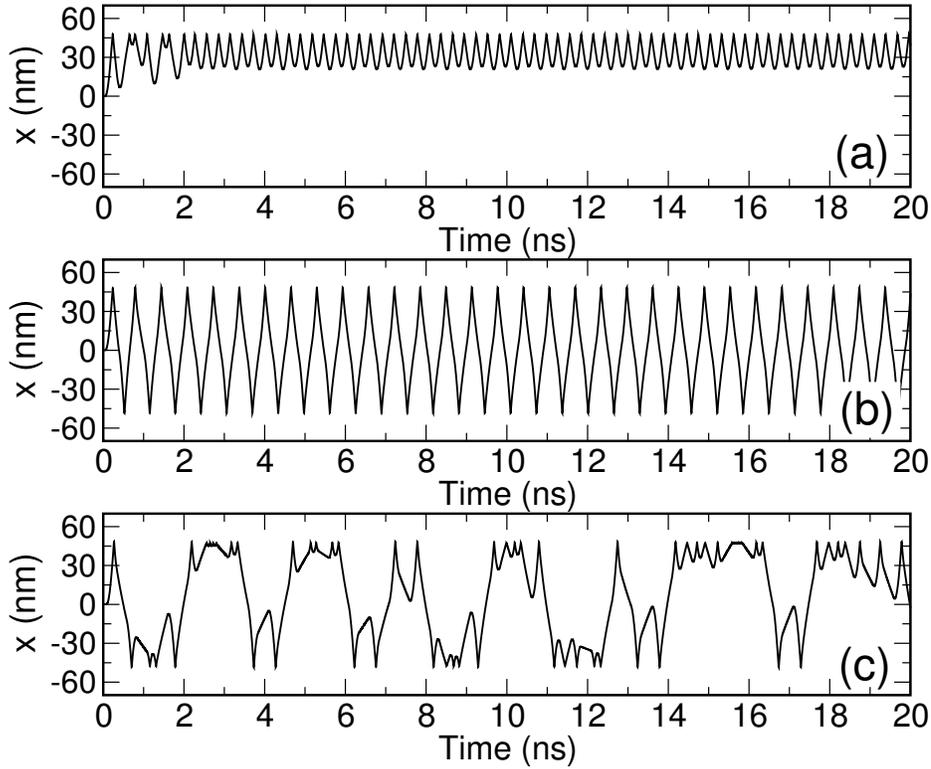}
\vspace{-0.4cm}
\caption{Time evolution of the inner tube obtained from Eq. (1) for three values of driven period: (a) 290 ps, (b) 320 ps, and (c) 500 ps. The data correspond for the same CNT oscillator and conditions of Fig. 2 (b). The oscillation frequency for the cases (a) and (b) are 3.4 and 1.6 GHz, respectively.}
\end{center}
\end{figure}

The behavior of the CNT-based oscillator subject to periodically driven thermal forces results from an interplay between the thermal gradient and the lengths of the inner and outer tubes. The inner tube length significantly affects the bifurcation diagrams and, consequently, the necessary driven periods for regular motions. The thermal gradient greatly influences the overall behavior of the CNT oscillator since, for relatively high values of $dT/dx$, the acceleration of the inner tube can be so high that it could be ejected, destroying the nanodevice. For long outer tubes ($\sim$ 1 $\mu$m), the inner tube can reach a steady velocity state which can change the system dynamics, not allowing periodic motions that cover all the outer tube (Fig.~4~(b)) \cite{mov}. From the practical point of view, very long outer tubes will be strongly influenced by off-axis tube oscillations. Preliminary MD simulations of models for the nanodevice shown in Fig.~1 indicated high amplitude off-axis oscillations of the outer tube when its extremities are not clamped. On the other hand, clamped extremities (e.g., achieved by the use of additional CNT junctions along the outer tube) can prevent such vibrations and increase the stability of the nanodevice proposed here.   


In summary, based on numerical calculations and molecular dynamics simulations we demonstrated that the use of thermal gradients can be an effective approach to initialize, control, and tune CNT-based oscillators. Financial support from the Brazilian agencies FAPESP (grant 2007/03923-1) and CNPq is acknowledged.

\clearpage

\begin{center}
{\Large\textbf{Supplementary Information}} \\
\textbf{Thermophoretically Driven Carbon Nanotube Oscillators}\\
V. R. Coluci, V. S. Tim\'oteo, and D. S. Galv\~ao
\end{center}

\section{Equation of Motion}

The motion of the inner tube of the CNT oscillator considering only translational motions (rigid tubes) is given by 
\begin{equation}
\displaystyle m\ddot{x} - F(x) + \gamma\sgn(\dot{x})|\dot{x}|^2 = \alpha \frac{dT}{dx}\; p(t)
\end{equation}
where the restoring force is written as

\begin{equation}
\displaystyle F(x)=F_{vdW}\left\{\Theta(x+r)[1-\Theta(x-r)][\Theta(-x-\Delta)-\Theta(x-\Delta)]\right\}
\end{equation}
with

\begin{equation}
\displaystyle \Theta(x)=\frac{1}{2}+\frac{x}{2\sqrt{x^2+\delta^2}}
\end{equation}
($\delta = 0.1 $ {\AA}) and
\begin{equation}
\displaystyle r\equiv\frac{L+l}{2} \;\;\;\;\Delta\equiv\frac{L-l}{2},
\end{equation}
where $F_{vdW}$ is the magnitude of the van der Waals restoring force, $L$ and $l$ are the outer and inner tube lengths, respectively. The form of $F(x)$ is shown in Fig. S1, for $L=100$ nm and $l=5$ nm. 

The thermal gradient was chosen to be practically constant on the outer tube length based on molecular dynamics simulations (see below) according to the following expression

\begin{equation}
\displaystyle \frac{dT}{dx}(x)\equiv \left| \frac{dT}{dx}\right| \left[\frac{2}{1+\exp(-30x)} - 1\right].
\end{equation}

Thus, the term $\alpha dT/dx$ represents the thermal force due to thermal vibrations of the outer tube caused by the externally imposed thermal gradient (Fig. S1). The regime of a constant thermal gradient is obtained after a transient period once the electrical current pulse is applied. We have analyzed the dynamics of the transient period using molecular dynamics simulations (see below).

The thermal gradient can be turned on or off accordingly to the presence or not of the electrical current flux through the supporting tube. The function $p(t)$ determines when the current would be on or off. We choose a square wave-like function to represent $p(t)$ with the form

\begin{equation}
\displaystyle p(t)= s(\bmod(t,P))
\end{equation}
where

\begin{equation}
\displaystyle s(\tau)\equiv \frac{1}{[1+\exp[-\beta(\tau-t_1)]]\;[1+\exp[-\beta(-\tau+t_2)]]},
\end{equation}
with $P$ being the period of $p(t)$. The modulo function $\bmod$ was used to provide the $p(t)$ periodicity. The parameters $\beta (=10 $ ps$^{-1})$, $t_1=P/6$, and $t_2=4t_1$ characterize $p(t)$. An example of $p(t)$ is depicted in Fig. 2 (B) for $P=400$ ps.

\begin{figure}[ht]
\begin{center}
\includegraphics[angle=0,scale=0.5]{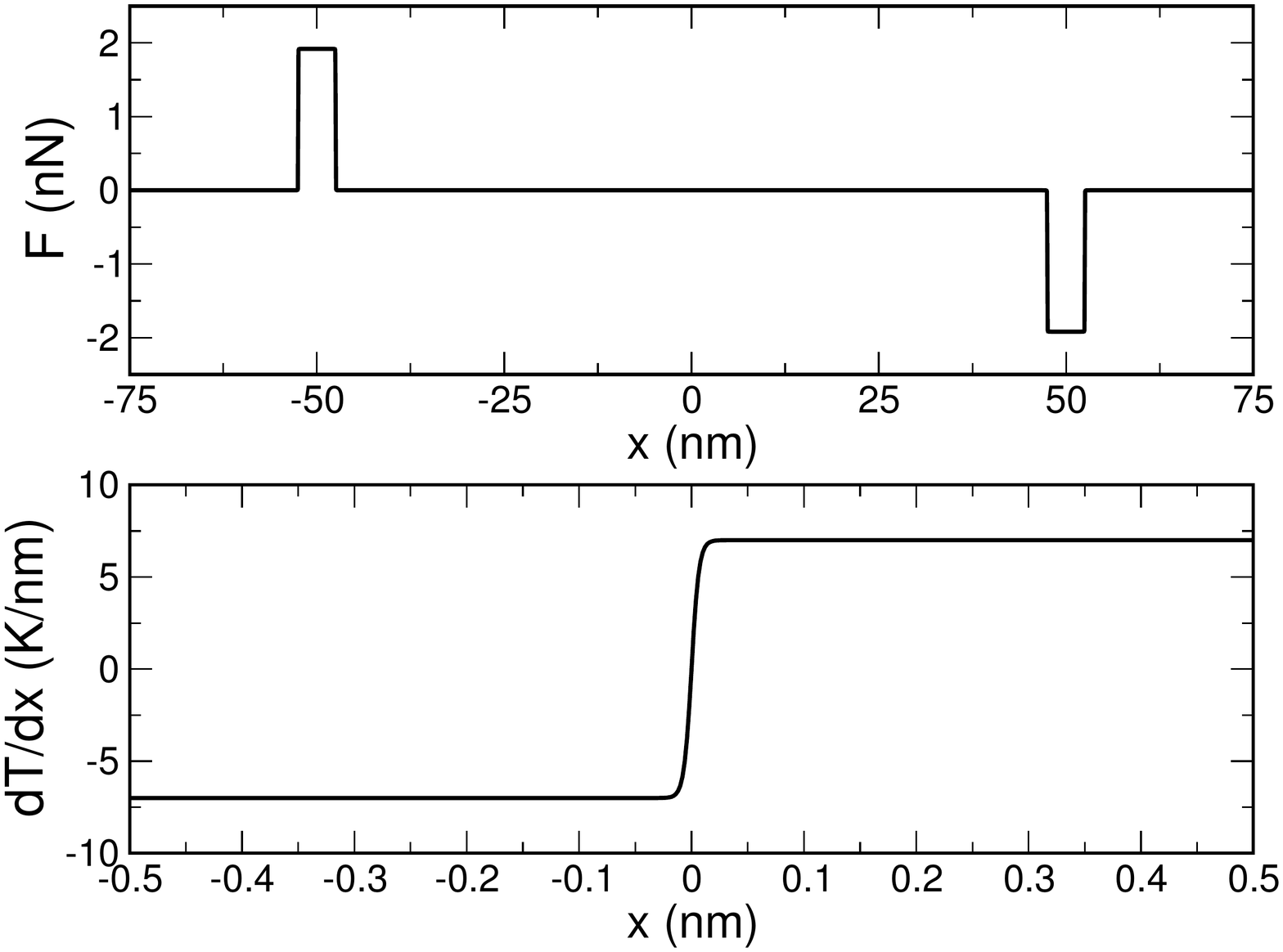}

FIG. S1: \footnotesize{$F(x)$ (Eq. 2) for $L=100$ nm, $l=5$ nm, and $F_{vdW}=1.92$ nN (top) and $dT/dx$ (Eq. 5) (bottom).}
\end{center}
\end{figure}

\section{Carbon oscillator model}

The model used in the molecular dynamics simulations is represented in Fig. S2. It is formed by a double-walled carbon nanotube with combination (10,10)@(15,15). Outer tube atoms that are fixed during the simulations are represented in blue, thermostated atoms in yellow ($T_1$) and red ($T_2$). Temperatures $T_1$ and $T_2$ were imposed at the beginning of the simulations.

\begin{figure}[h]
\begin{center}
\includegraphics[angle=0,scale=0.6]{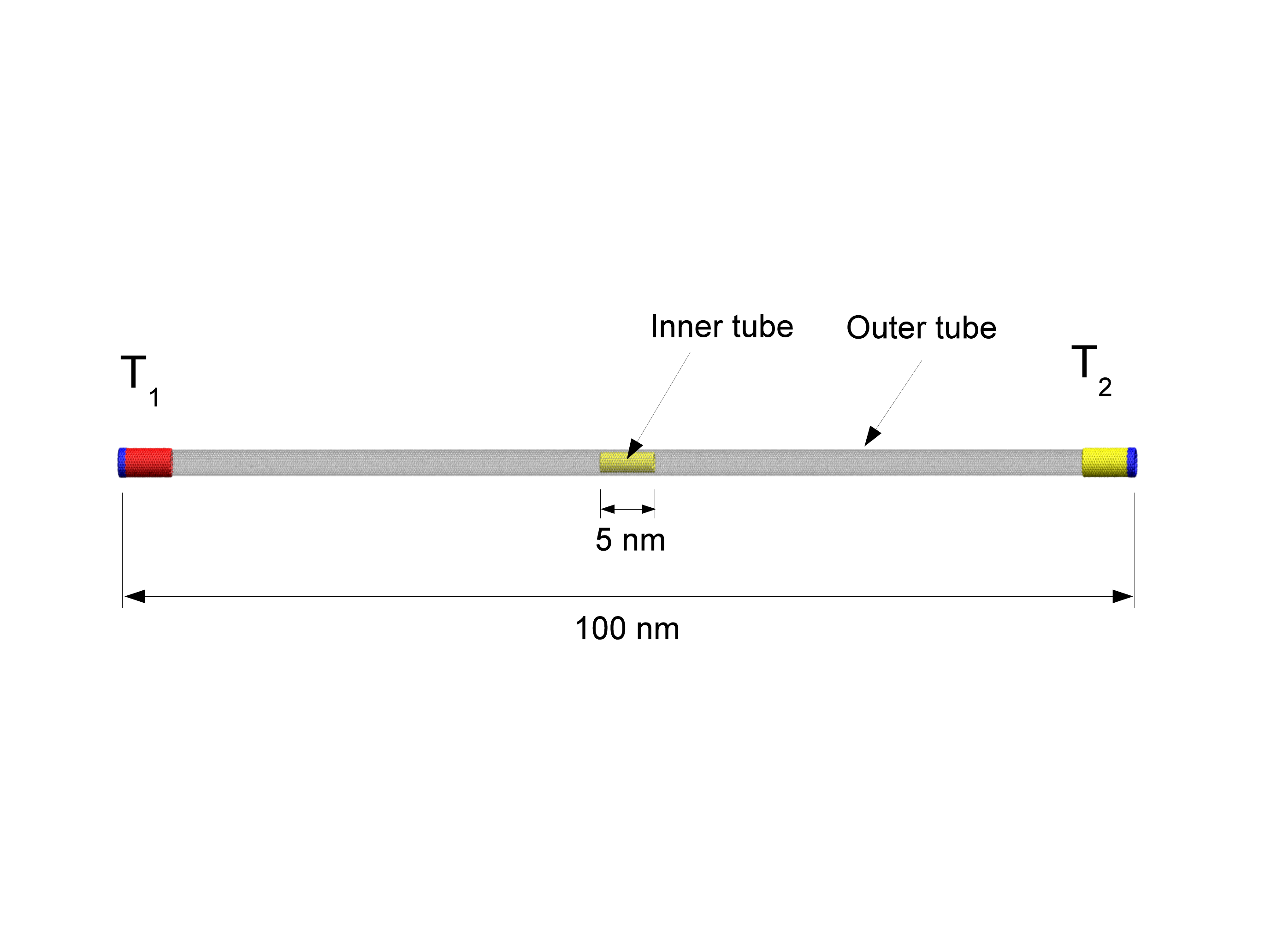}

FIG. S2: \footnotesize{Model used in the molecular dynamics simulations.}

\end{center}
\end{figure}

We used this model to determine the temperature profile along the outer tube as a function of time once the thermal gradient is created. The analysis of the movement of the inner tube motion provided the values for the parameters $F_{vdw}$, $|dT/dx|$, $\gamma$, and $\alpha$.

\section{Temperature Profile}  

Fig. S3 presents the temperature profile of the outer tube for the system (10,10)@(15,15) for different times after a thermal gradient of 7 K/nm is imposed between the two extremities of the outer tube. As we can see an approximately linear variation of the temperature with $x$ is attained after about 50 ps, the period $P$ of $p(t)$ should be larger than 100 ps in order to keep the CNT oscillator operating in the constant thermal gradient regime. 

\begin{figure}[ht]
\begin{center}
\includegraphics[angle=0,scale=0.4]{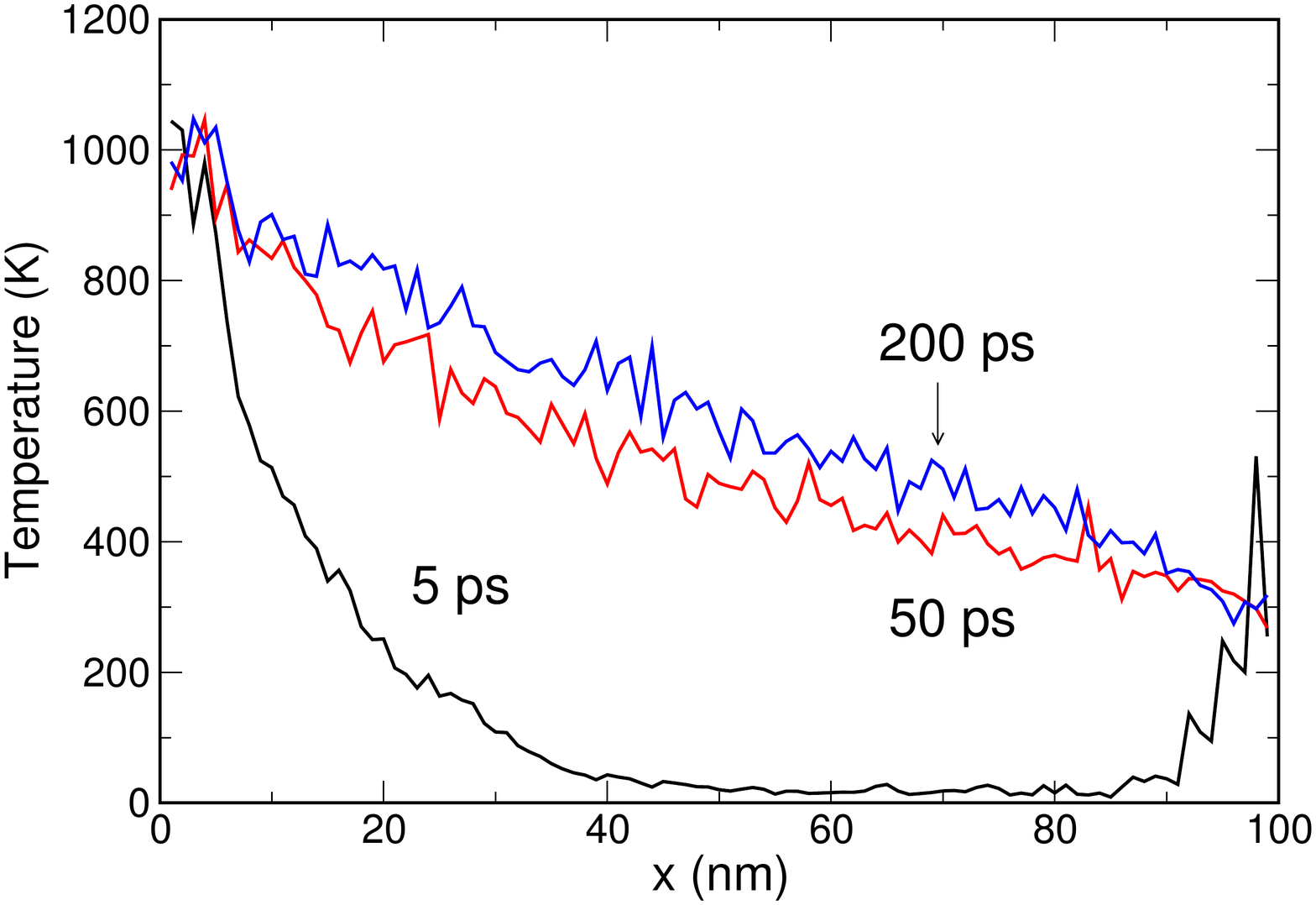}

FIG. S3: \footnotesize{Temperature profile of the outer tube (15,15) for the model of Fig. S2 at different times of the simulation. The temperature was obtained from the kinetic energy of atoms averaged over equally spaced regions of 1 nm.}

\end{center}
\end{figure}

\section{Dependence upon tube lengths and thermal gradients}

As mentioned in the main text the behavior of the thermophoretically driven CNT oscillator depends upon both the inner tube length and the thermal gradient. In order to evaluate this dependence, we identified  the cases where there is an oscillation (regular or chaotic) of the inner tube. In such cases the motion is limited within the outer tube. On the other hand, for some set of parameters, the inner tube was strongly accelerated and ejected off the outer tube. The evolution of the inner tube position was obtained by numerical integration of Eq. (1) for different values of the thermal gradient and driving period $P$. These results are only estimations since we used parameters for the case of inner tube length of 5 nm and thermal gradient of 7 K/nm ($F_{vdW}= 1.92 $ nN, $\gamma = 1.48 $ pN ps$^2$/{\AA}$^2$, and $\alpha = 8.08 $ pN nm/K) to all other thermal gradient cases.

Figure S4 shows a phase diagram which indicates, for which combinations of the driving period and the thermal gradient $|dT/dx|$, the inner tube motion is limited. As one can see, the phase diagram exhibits a complicated structure. The longer the inner tube, the higher the thermal gradient necessary to completely eject the inner tube.

\begin{figure}
\begin{center}
\includegraphics[angle=0,scale=0.25]{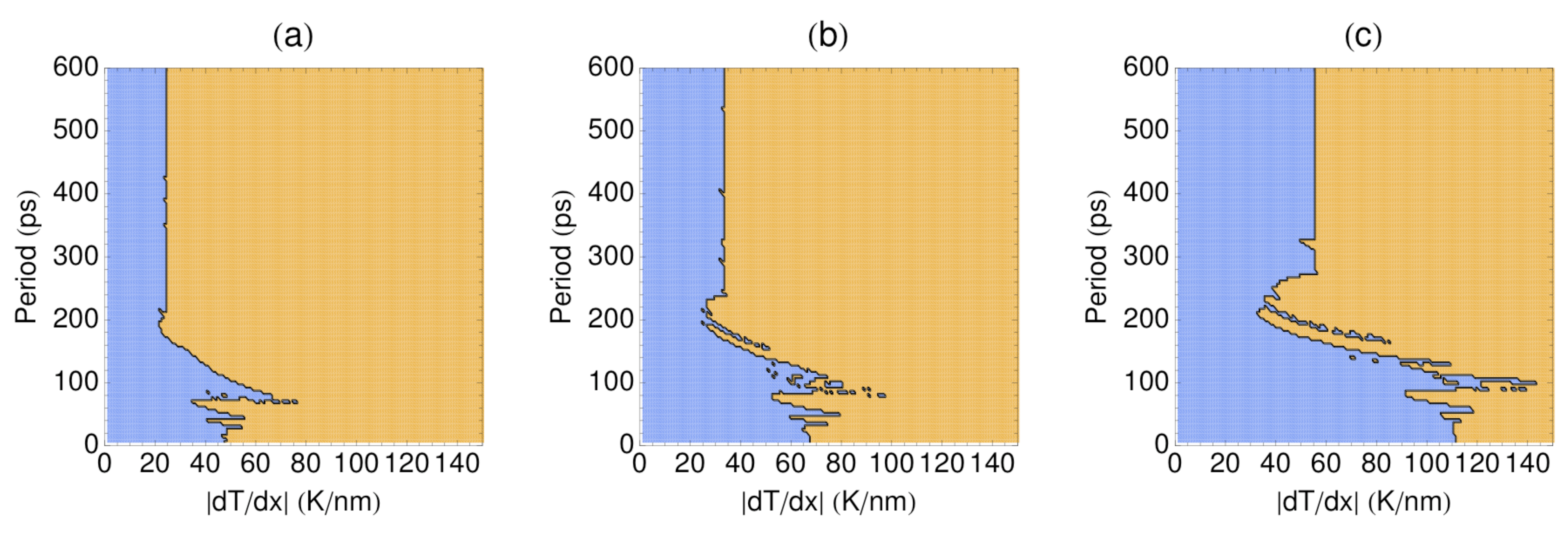}

FIG. S4: \footnotesize{Phase diagram for the driving period $P$ and the thermal gradient $|dT/dx|$ for (10,10)@(15,15) CNT-based oscillator with inner tube lengths of (a) 5 nm, (b) 10 nm, and (c) 20 nm (outer tube length of 100 nm). Blue region indicates the combination of period and thermal gradient which leads to a limited motion of the inner tube. Brown region indicates combinations where the inner tube is ejected from the outer tube.}

\end{center}
\end{figure}

The overall behavior of the oscillator will also depend upon the outer tube length. The inner tube can reach a steady velocity state if the outer tube is long enough. For example, for the thermal gradient of 7 K/nm and an inner tube length of 5 nm, steady velocity state was observed for an outer tube length of 1 $\mu$m, as shown in Figure S5. For this case, the behavior of the device will be different as represented in the bifurcation diagram of Figure S6. The inner tube can now perform periodic motions only at one side of the outer tube. This is because the inner tube is not able to return to the middle of the outer tube for such large outer tube length. Furthermore, such large outer tube lengths may cause additional issues such as off-axis vibrations (if not clamped) which may make the device not viable. Modifying the thermal pulse (form, periods, etc) might provide a way to allow periodic motions covering all long outer tubes.

\begin{figure}
\begin{center}
\includegraphics[angle=0,scale=0.5]{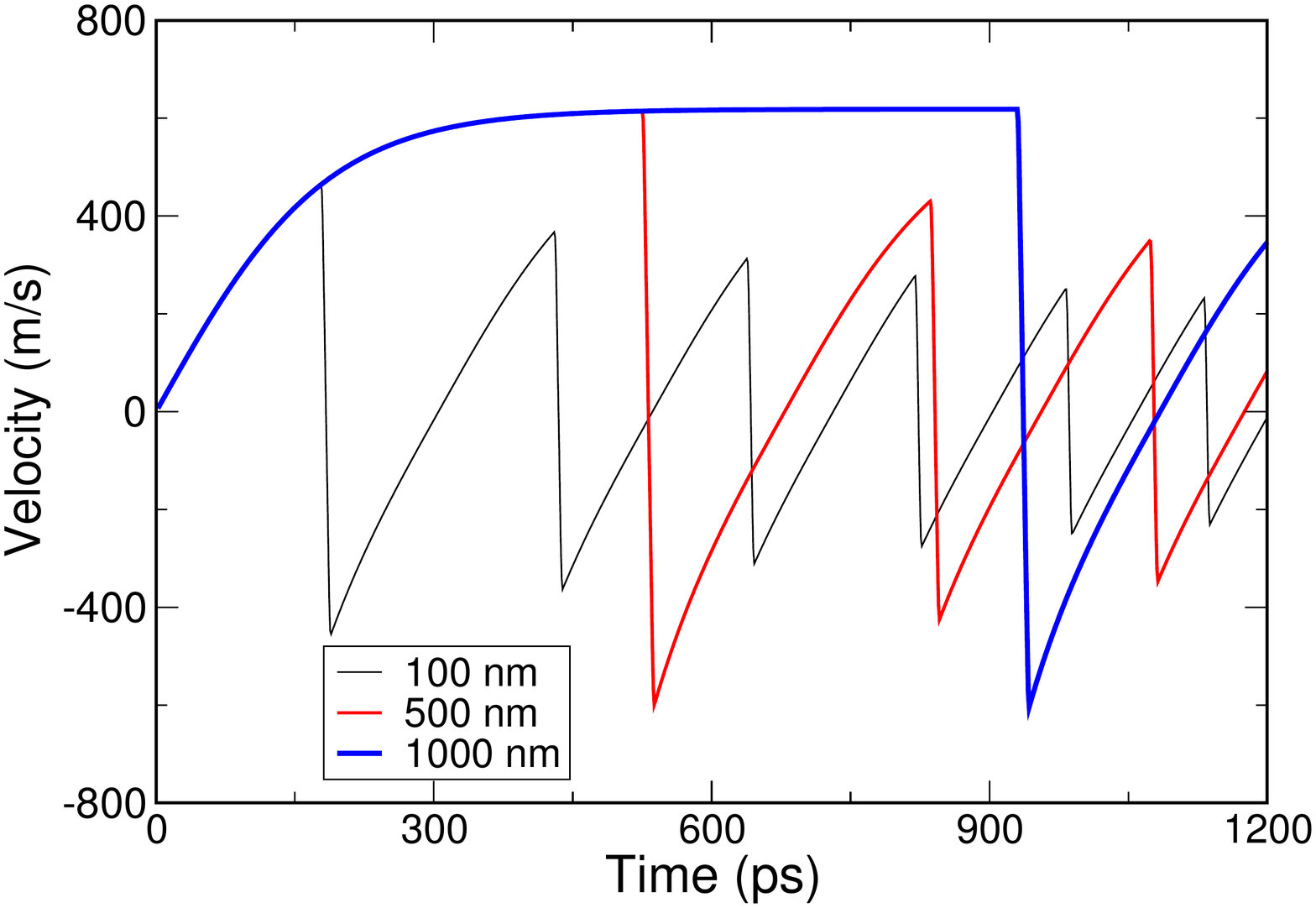}

FIG. S5: \footnotesize{Time evolution of the velocity of the inner tube center of mass obtained from integration of Eq. (1) for different outer tube lengths ($p(t)$ = 1). The data correspond to the following parameters: $F_{vdw}=$ 1.92 nN, $\alpha = $ 8.08 pN nm/K, $\gamma = 1.48 $ pN ps$^2$/{\AA}$^2$, thermal gradient of 7 K/nm, and inner tube length of 5 nm.}

\end{center}
\end{figure}

\begin{figure}
\begin{center}
\includegraphics[angle=0,scale=0.3]{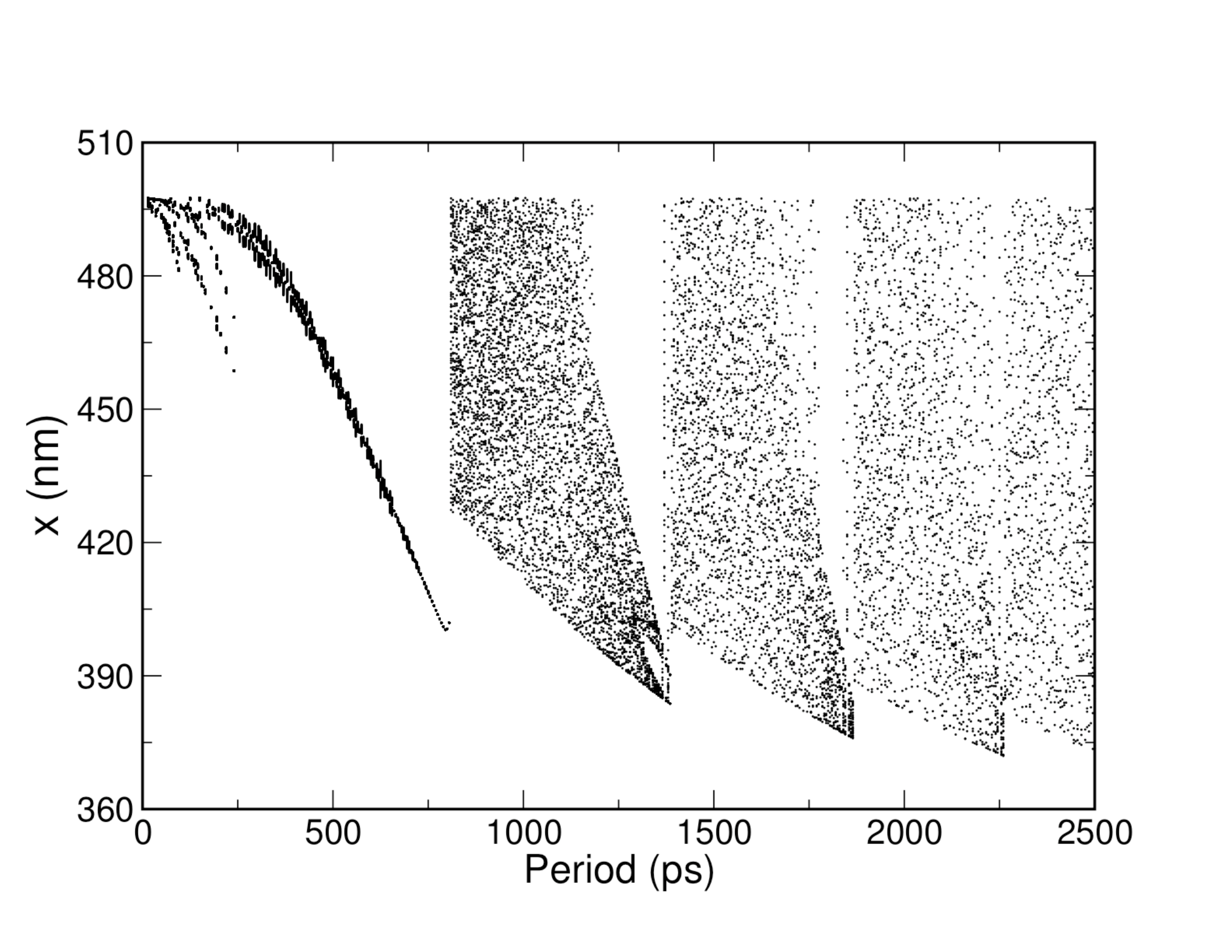}

FIG. S6: \footnotesize{Bifurcation diagram for $x$ as a function of the period of $p(t)$ for a thermal gradient of 7 K/nm. The data correspond to the (10,10)@(15,15) with inner tube and outer tube lengths of 5 nm and 1000 nm, respectively, for the following parameters: $F_{vdw}=$ 1.92 nN, $\alpha = $ 8.08 pN nm/K, and $\gamma =$ 1.48 pN ps$^2$/{\AA}$^2$.}

\end{center}
\end{figure}

For low thermal gradients (below $\sim$ 1 K/nm), steady velocity state is expected to be achieved for even longer outer tubes. By reducing the temperature, a decreased in the friction is expected since nanotube vibrations (one of the causes of energy dissipation) are reduced. This is indeed observed in the $\gamma$ values of the fitted parameters for the cases of 7 and 17 K/nm. Thus, we expect a decreasing of the $\gamma$ value for lower thermal gradients than 7 K/nm.

Since we have not performed molecular dynamics simulations for low thermal gradients (below 1 K/nm), which provide the necessary parameters for Eq. (1), we can only estimate the behavior for such cases. The estimation is done by numerically integration Eq. (1) for different parameters of $\gamma$ and $\alpha$. The parameter $F_{vdw}$ will not change since we are not changing the double-walled carbon nanotube considered.

Keeping the same $\alpha$ value of the 7 K/nm case and using $\gamma = $ 0.5 pN ps$^2$/{\AA}$^2$, we can see in Figure S7 that the steady velocity state will be achieved for even longer outer tubes for a thermal gradient of 0.1 K/nm. Thus, the value of 1 $\mu$m for the outer tube length can be considered as a lower limit where stead velocity state is expected, for thermal gradients below $\sim$ 10 K/nm.

\begin{figure}
\begin{center}
\includegraphics[angle=0,scale=0.5]{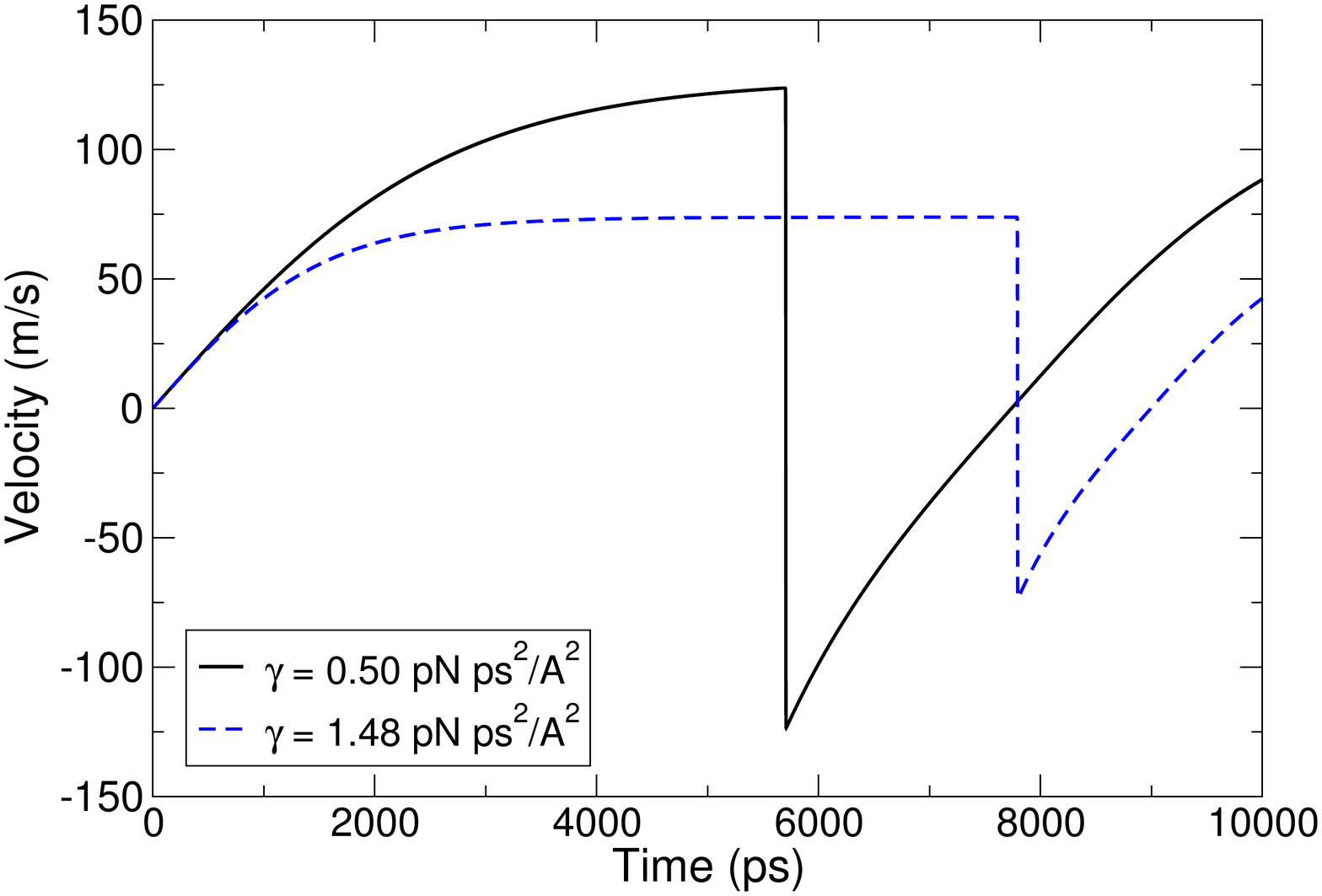}

FIG. S7: \footnotesize{Time evolution of the velocity of the inner tube center of mass obtained from integration of Eq. (1) with the following parameters: $F_{vdw}=$ 1.92 nN, $\alpha = $ 8.08 pN nm/K, and a thermal gradient of 0.1 K/nm ($p(t)$ = 1). The inner and outer tube lengths were 5 nm and 1000 nm, respectively.}

\end{center}
\end{figure}

\end{document}